\documentclass{pasj00} 
\begin{document}
\SetRunningHead{Y. Sofue}{Optimizing Rotation Curve Observations}
\Received{2008/mm/dd}  \Accepted{2008/mm/dd} 

\def\kms{km s$^{-1}$}  \def\Msun{M_\odot} 
\def\be{\begin{equation}} \def\ee{\end{equation}}
\def\bc{\begin{center}} \def\ec{\end{center}}
\def\Rsun{R_0} \def\Vsun{V_0} \def\sin{{\rm ~sin~}} \def\cos{{\rm ~cos~}}
\def\vr{v_{\rm r}} 
\def\vp{v_{\rm p}} 
\def\p{\pm}
\def\sinl{~{\rm sin~}l~} 
\def\cosl{~{\rm cos~}l~}
\def\masy{mas y$^{-1}$}
\def\Vrotr{V_{\rm rot}^{v_{\rm r}}}
\def\Vrotp{V_{\rm rot}^\mu}
\def\Vrotv{V_{\rm rot}^{\rm vec}}
\def\vr{v_{\rm r}}
\def\Up{U_{\rm p}}
\def\Ur{U_{\rm r}}
\def\V0{V_0}
 \def\p{$\pm$}

\title{Accuracy Diagrams for the Galactic Rotation Curve and Kinematical Distances}
\author{Yoshiaki {\sc Sofue}}  
\affil{Department of Physics, Meisei University, Hodokubo 2-1-1, Hino, Tokyo 191-8506\\
and\\
Institute of Astronomy, The University of Tokyo, Osawa 2-21-1, Mitaka, Tokyo 181-0015 \\
sofue@ioa.s.u-tokyo.ac.jp}

\KeyWords{galaxies: the Galaxy --- galaxies: ISM --- galaxies: kinematics --- galaxies: rotation curve --- astrometry: proper motion --- astrometry: distance} 

\maketitle

\begin{abstract}  
We revisit the methods to determine the Galactic rotation curve and kinematical distances from radial velocities and proper motions. We construct "accuracy diagrams" to show the distributions in the galactic plane of expected uncertainties in the derived quantities such as rotation velocities and kinematical distances. We discuss how to optimize the source selection for measurements of kinematical quantities based on the accuracy diagrams.
\end{abstract}

\section{Introduction}

Rotation curve is the major tool to study the dynamics and structure of the Galaxy (Binney and Merrifield 1998; Sofue and Rubin 2001; Sofue et al. 2009). Once it is determined, the rotation curve is used to obtain the distribution of mass in the Galaxy, and map the interstellar matter from kinematical observables. Accuracy of the obtained results depends not only on the observational accuracy, but also on the methods as well as on the location of observed objects in the galactic disk. However, accuracy analyses have been not systematically obtained in the current studies of galactic kinematics.

VLBI measurements of parallaxes and proper motions of galactic maser sources have opened a new era in the study of galactic kinematics, particularly in determining the rotation curve and distances of objects (e.g. Honma et al. 2007; Reid et al. 2009). VERA has played the essential role, and high accuracy determination of the rotation velocity is now available using trigonometric distance measurements combining with proper-motion (Honma et al. 2007).

A more general method to measure the distance, radial velocity, and proper motion at the same time has been applied for determining three-dimensional velocity vectors of maser sources (Oh et al. 2010). These measurements are expected to provide us with a global rotation curve constructed from objects distributed over the galactic disk. This will improve the accuracy not only of the outer rotation curve, but also of the inner rotation curve, where the current observations have been obtained mostly from the tangent point data.

Accordingly, observations of a larger number of objects distributed in the galactic disk have become possible for deriving the rotation curve. We remember that the accuracy of derived rotation velocity depends not only on the intrinsic observing errors, but also on the location of measured objects. 
 
In this paper, we analyze the behaviors of observable kinematical quantities such as the radial velocity and proper motions in the galactic disk. We investigate the dependence of the accuracy of derived rotation velocities on the galactic positions of the observed sources. The results would be useful as a guide for optimizing the selection of objects for determination of the Galactic rotation curve. The present analyses will be given on the assumption of circular rotation of the galactic objects. We present the result only for a fixed set of the galactic constants, $R_0=8$ kpc and $V_0=200$ \kms. However, the analysis does not include systematic errors arising from the uncertainties of these quantities. Hence, this paper should be taken as a methodological guide, and a more practical use may be recalculated for such constants in individual cases and data sets.

\section{Rotation Curve}

 The tangent-point method, or the terminal velocity method, has been most often applied to determine the inner rotation curve (Clemens 1985). The outer rotation curve has been determined using spectro-photometric distances combined with radial velocities of interstellar lines (Fich et al. 1989), or by the HI disk thickness method (Merrifield 1991; Honma and Sofue 1997). However, the outer rotation curve is still crude mainly because of the distance uncertainties.

\begin{figure} 
\bc
\includegraphics[width=8.5cm]{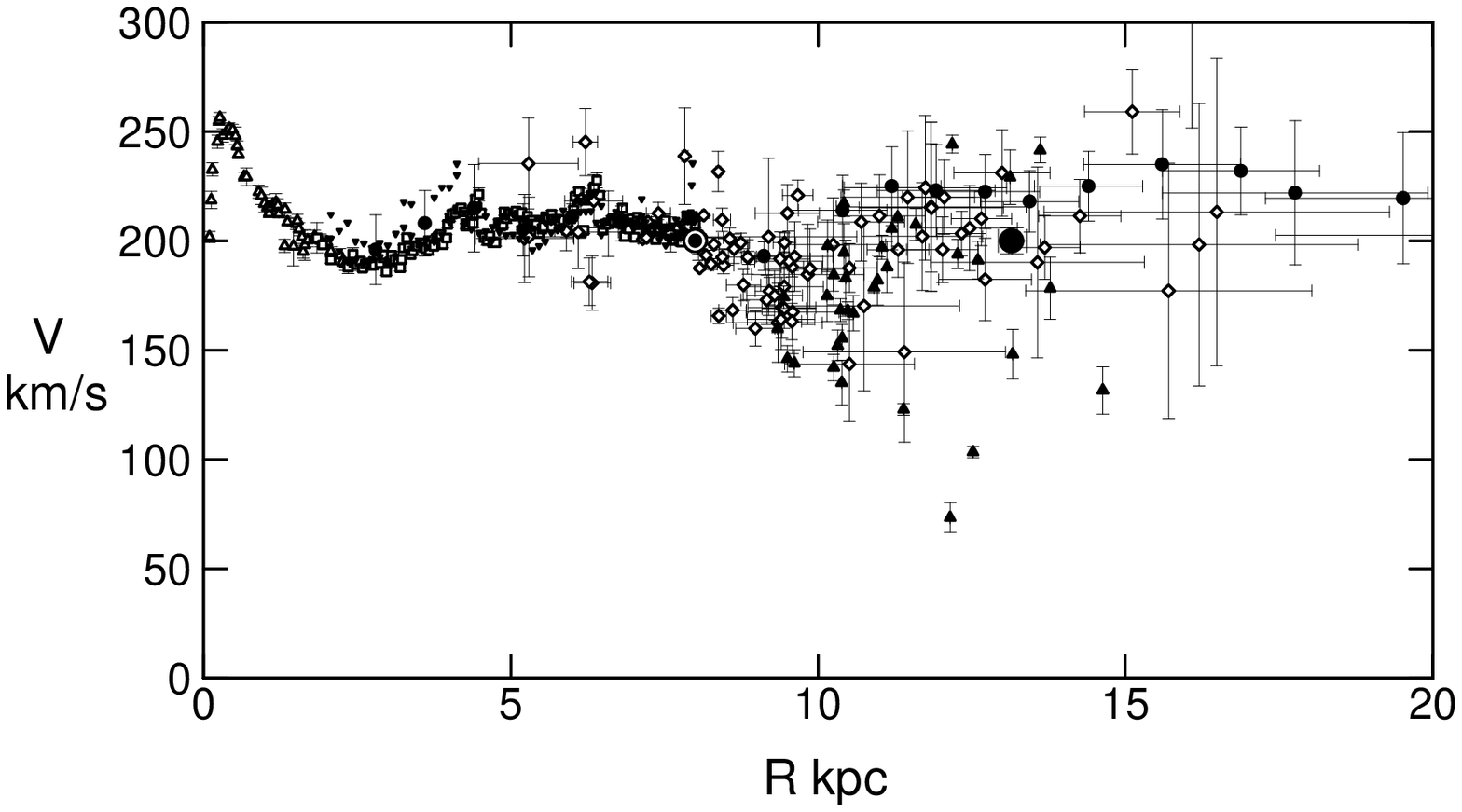}  \\  
\ec
\caption{Rotation curve of the Galaxy (Sofue et al. 2009). Inner curve at $r<8$ kpc is obtained mainly by the tangent-point method,  while the outer curve is crude because of the ambiguities in distance estimation. Big diamonds are from recent VERA observations (see table \ref{tab-veravrot}).}
\label{fig-rc}  
\bc
\includegraphics[width=8cm]{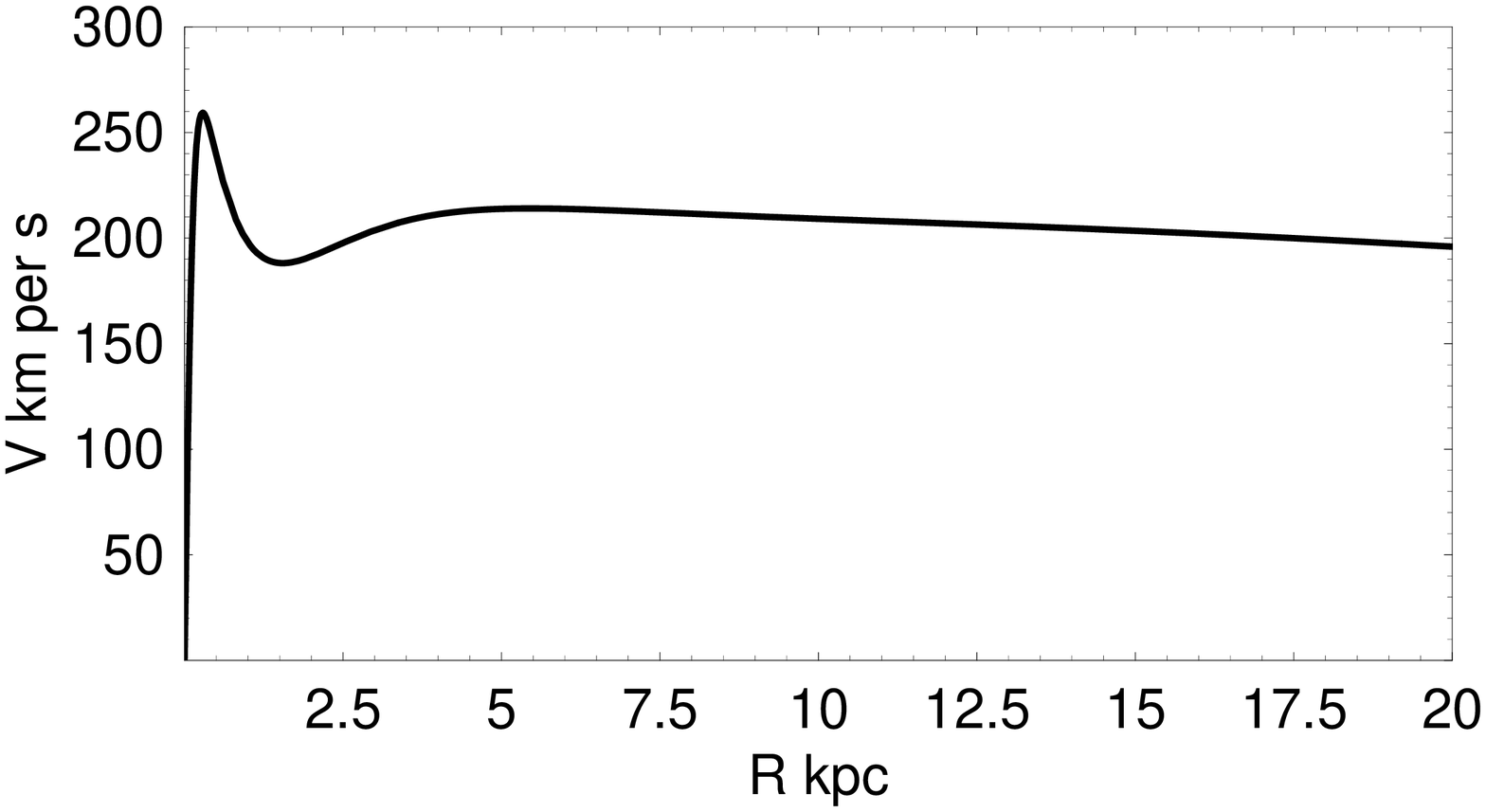}  
\ec
\caption{A Plummer model rotation curve $V(R)$ for parameters in table \ref{tab-plummer}.}
\label{fig-rcmodel}  
\end{figure}

 Figure \ref{fig-rc} shows a rotation curve of the Galaxy obtained by compiling measured rotation velocities from the literature in the decades (Sofue et al. 2009). The galactic constants are taken to be $R_0=8.0$ kpc and $V_0=200$ \kms in this paper. Recent values from VERA observations are also included (see table \ref{tab-veravrot}). Rotation velocities within the solar circle are accurately determined using the tangent point method, which uses the terminal velocities in the spectral lines of interstellar gases (Clemens 1985). Since this method does not require distances of the emission regions,  it gives relatively high accuracy of rotation velocity. On the other hand, the outer rotation curve is crude, because the distance measurements of objects, which usually contain large errors, are inevitable (Blitz 1979; Fich et al. 1989; Merrifield 1992; Honma and Sofue 1997; Binney and Dehnen 1997).

In order to discuss the accuracy of derived rotation velocities in the following sections, we need to use an approximate rotation curve to represent the observations. Since the present study treats the kinematics only in the galactic plane, we here adopt a simple model of three-component Plummer potential.
\be
V(R)=\sqrt{R {\partial \Phi \over \partial R}},
\label{eq-vrot}
\ee
and 
\be
\Phi=\Sigma {G M_i \over \sqrt{R^2+a_i^2}},
\label{eq-phi}
\ee
where  $G$ is the gravitational constant, $a_i$ are the scale radii of the individual mass components, and $M_i$ are the masses. Table \ref{tab-plummer} lists the values of the parameters, and figure \ref{fig-rcmodel} shows the model rotation curve, which approximately represents the observations in figure \ref{fig-rc} with $R_0=8$ kpc and $V_0=200$ \kms. We note that the calculated results in the next sections are not strongly dependent on the shape of rotation curve.

\begin{table}
\caption{Plummer potential parameters for a model galactic rotation curve, mimicking the observed rotation curve in figure \ref{fig-rc}. }
\begin{tabular}{lllllll}
\hline\hline 
Component $i$ & $a_i$ (kpc) & $M_i~(\Msun)$ \\
\hline
1 & 0.2 & $0.8\times 10^{10}$ \\
2 & 3.6 & $0.6 \times 10^{11}$ \\
3 & 15  & $2.0 \times 10^{11}$ \\
\hline  \\
\end{tabular}
\label{tab-plummer}
\end{table}

\begin{figure} 
\bc
\includegraphics[width=7cm]{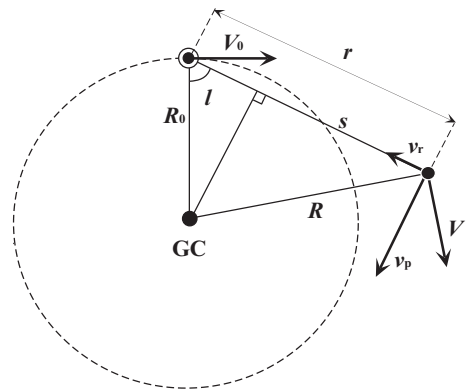}  
\ec
\caption{Definition of used variables and parameters.}
\label{fig-circ}  
\end{figure}

\section{Accuracy Diagrams for Rotation Curve}

We denote the radial velocity by $\vr$, and perpendicular velocity to the line of sight by $\vp=\mu r$ with $\mu$ being the proper motion and $r$ the distance to the object from the Sun. These quantities are related to the circular rotation velocity $V$  as
\be
\vr= \left({R_0 \over R} V - V_0\right) \sinl,
\label{eq-vr}
\ee
and
\be
\vp=\mu r =-{s \over R} V -V_0 \cosl,
\label{eq-vp}
\ee
where
\be
s=r-R_0 \cosl
\ee
Here $R$ is the galacto-centric distance, and is related to $r$ and galactic longitude $l$ as 
\be
R=\sqrt{r^2+R_0^2 - 2 r R_0 \cosl}.
\ee
Figure \ref{fig-circ} illustrates the definition of used variables and parameters in this article.

\subsection{Rotation Velocity $\Vrotr$ from Radial-Velocity, and Accuracy Diagram, $\Delta \Vrotr(X,Y)$} 

If we assume that the object's orbit is circular around the Galactic Center, the rotation velocity $V$ can be obtained by measuring the radial velocity $v_{\rm r}$ and its distance $r$, which is expressed by the galacto-centric distance $R$ and longitude $l$:
\be
\Vrotr= {R \over R_0} \left({v_{\rm r} \over \sinl} +V_0 \right).
\label{eq-V}
\ee
Since the observations includes errors in $\vr$ and $r$, the resultant rotation velocity $V$ has an error which is expressed by
\be
\Delta \Vrotr=\sqrt{\delta V_{\rm vr}^2+\delta V_{\rm r}^2}.
\ee
Here,
\be
\delta V_{\rm vr}={\partial V \over \partial v_{\rm r}} \delta v_{\rm r}, ~~~~
\delta {V_{\rm r}}={\partial V \over \partial r} \delta r.
\ee
We obtain
\be
\Delta \Vrotr 
=\left[{\left(R \over R_0 \sinl \right)^2\delta v_{\rm r}^2
+ \left(s ~V \over R^2 \right)^2 \delta r^2}\right]^{1/2}.
\label{eq-dvrad}
\ee
The uncertainty in the galacto-centric distance $R$ arises from the error in distance measurement as 
\be
\delta R = {s \over R} \delta r.
\label{eq-deltaR}
\ee
 Note that
\be
\sinl=X/r,~~~ \cosl=-(Y-R_0)/r
\ee
in the Cartesian coordinates centered on the Galactic Center. Since equation \ref{eq-dvrad} includes the rotation velocity $V$, the error distribution depends on the rotation curve. 

\begin{figure} 
\bc
\includegraphics[width=7.5cm]{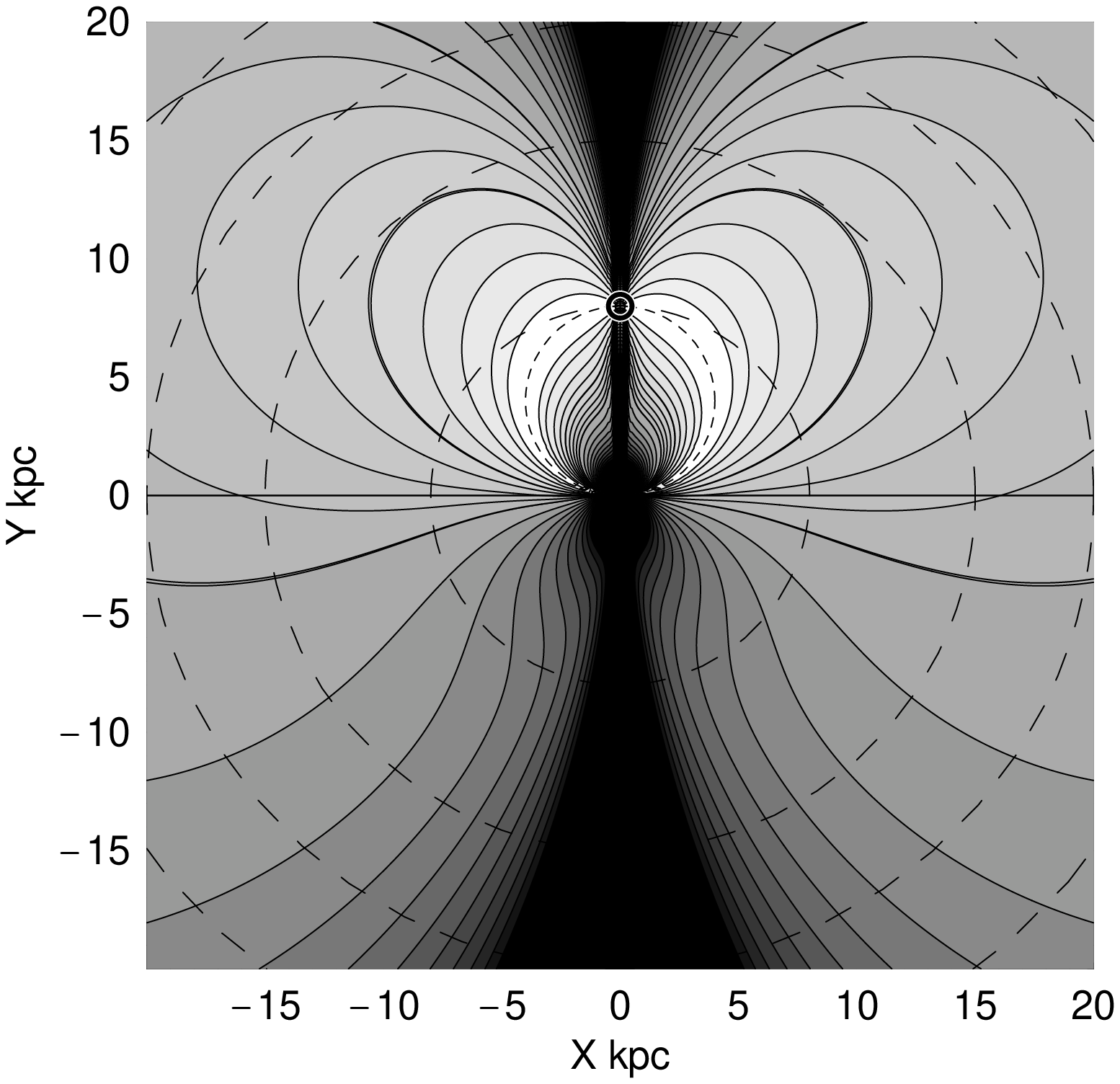}  \\  
\ec
\caption{Accuracy diagram $\Delta \Vrotr(X,Y)$ for $\delta \vr=1$ \kms, $\delta r/r=0.02 $ (2\% distance error). Dashed circles represent $R=8$ kpc (solar circle), $R=15$, 20 and 25 kpc.  Contours are drawn from white to black at 1.2, 1.5, 2, 2.5, ..., 5, 6, 7, .... with 3 and 5 \kms by thick lines. Sources near the Sun-GC line yield the largest error. }
\label{fig-vrad}  
\bc
\includegraphics[width=7.5cm]{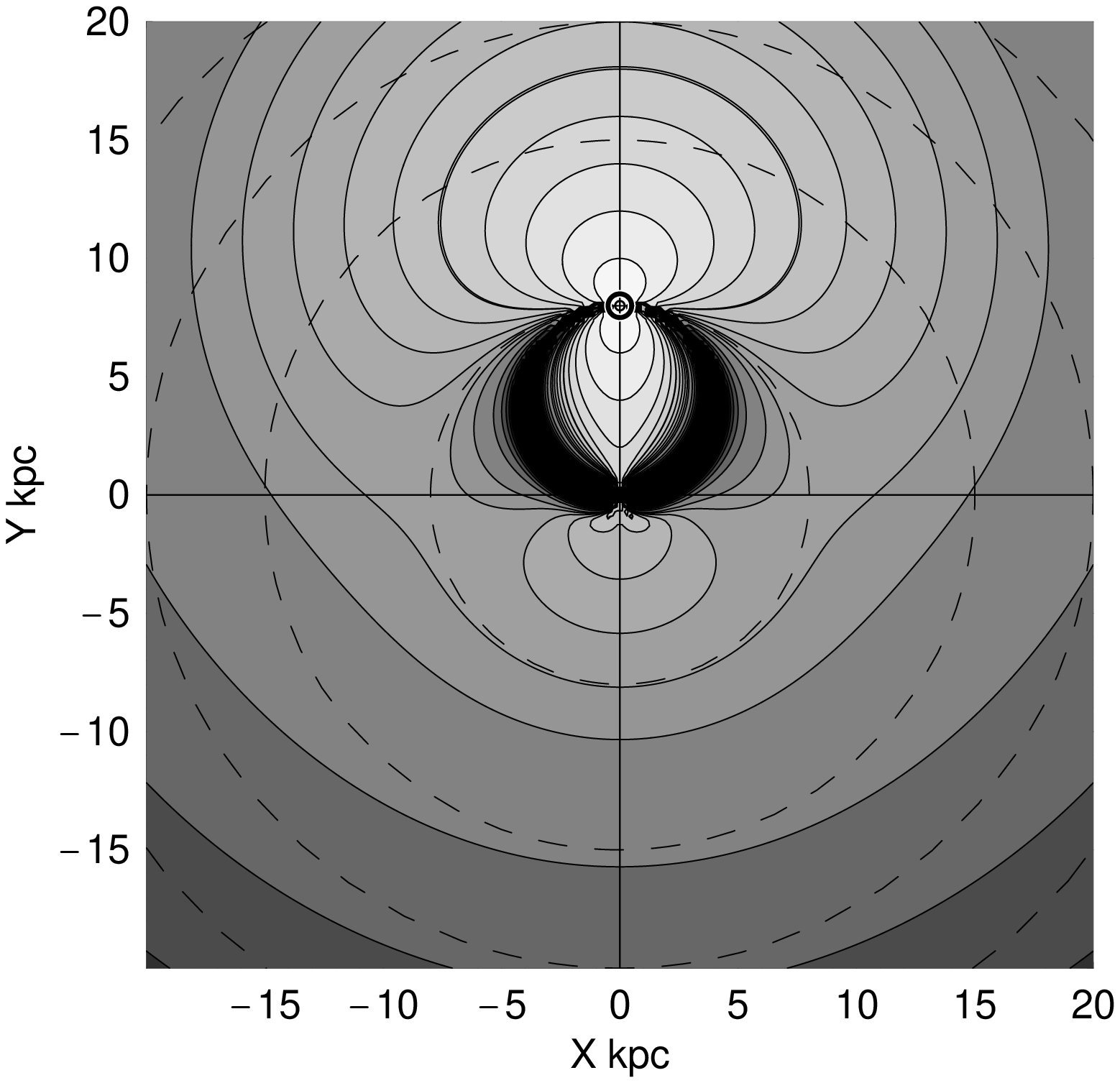}  \\  
\ec 
\caption{Accuracy diagram $\Delta \Vrotp(X,Y)$ for  $\delta \mu= 0.21 {\rm mas~y^{-1}} $  and $\delta r/r=0.02$ for 2\% distance error. Contours are drawn at 2, 4, ..., 20, 25, 30, ... \kms with 10 \kms by thick line. The error becomes largest along the tangent point circle.}
\label{fig-vprop}  
\end{figure}

Figure \ref{fig-vrad}  shows the thus calculated distribution of the expected error in rotation velocity, $\Delta \Vrotr$, by a contour map in the Cartesian coordinates $(X,Y)$. We may call this diagram the "accuracy diagram" for the rotation velocity. The calculation was made for a combination of $\delta \vr=1$ \kms and $\delta r/r=0.02 $, or 2\% error in distance measurement.  The regions with higher accuracy or with smaller errors are presented by bright area, while regions with larger errors are dark.  

This figure indicates that the accuracy is highest along the tangent point circle. Along this circle $s= r - R_0 \cosl =0$, and the second term in equation \ref{eq-dvrad} is equal to zero. 
For different parameters, the diagram may change quantitatively, but the overall characteristics remain unchanged, and hence, this diagram represents the general behavior of the accuracy distribution.

This is obviously the reason why the tangent-point method has resulted in higher-accuracy rotation curve inside the solar circle as in figure \ref{fig-rc}. Thus, the tangent-point circle is a special region for accurate rotation curve determination from radial velocity observations. Outside the tangent-point circle, the error is smoothly minimized in broad "butterfly" regions around $l\sim 100-135\deg$ and $l\sim 225-280\deg$.

On the other hand, this method yields the largest error near the Sun-GC line, where the direction of the circular rotation is perpendicular to the line-of-sight velocity, so that small observational error in the radial velocity largely affects the resultant rotating velocity. The Sun-GC line is, thus, the singular line in this method.

The uncertainty in $R$ is calculated as in equation (\ref{eq-deltaR}), and propagates to the uncertainty in $\Vrotr$ as equation (\ref{eq-dvrad}).  Therefore, equation (\ref{eq-dvrad}) already includes the uncertainty  $\delta R$ caused by $\delta r$. The uncertainty in $R$ also causes the uncertainty in model $V(R)$. This affects the result by the second order smallness, because the rotation curve is assumed to be nearly flat in most regions. However, $\delta R$ becomes very large at small $R$, as equation (\ref{eq-deltaR}) indicates. Therefore, the accuracy diagram should not be taken serious near the Galactic Center. These arguments apply similarly to the following sections.

\begin{figure}
\bc
\includegraphics[width=7.5cm]{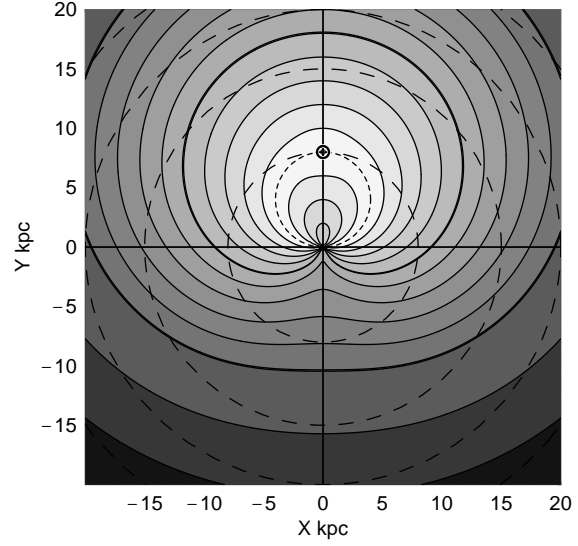}  \\   
\ec 
\caption{Accuracy diagram $\Delta \Vrotv(X,Y)$ for  $\delta \mu=0.21$ mas y$^{-1}$, $\delta \vr =1$ \kms, and $ \delta r/r=0.02$.. Contours are drawn at 2, 4, 6, .... 20, 25, 30, ... \kms with 10 and 20 \kms by thick lines. }
\label{fig-vvec}  
\end{figure}

\subsection{Rotation Velocity $\Vrotp$ from Proper Motion, and Accuracy Diagram, $\Delta \Vrotp(X,Y)$} 

If we assume circular motion, the rotation velocity is also determined by measuring the proper motion $\mu$ as
\be
\Vrotp = - {R \over s} (r \mu + V_0 \cosl).
\label{eq-vrotprop}
\ee 
In the same way as in the previous section and remembering that $R^2-s^2=R_0^2{\rm sin}^2 l$, we have 
\be
\Delta \Vrotp = \sqrt{\left(\partial V \over \partial r \right)^2 \delta r^2 + \left(\partial V \over \partial \mu \right)^2 \delta \mu^2}
\ee
\be
={R \over |s| }  \left[ r^2 \delta \mu^2 +  \left( {R_0^2  V {\rm sin}^2 l \over s R^3} + {\vp \over r}   \right)^2 \delta r ^2 \right]^{1/2}.
\label{eq-dvprop}
\ee 
The errors in  $v_{\rm p}$ and $r$ may be assumed to be proportional to the distance. We here calculate an accuracy diagram for $\delta \mu=0.21 $mas y$^{-1}$, and  $\delta r/r= 0.02$. Figure \ref{fig-vprop} shows the thus calculated accuracy diagram $\Delta \Vrotp$  for the same rotation curve as in figure \ref{fig-rcmodel}.

Figure \ref{fig-vprop} shows that the error becomes smallest along the Sun-GC line, and is still small in a large area in the anti-center direction. On the other hand, the error is largest around the tangent point circle, and the error equation \ref{eq-dvprop} diverges on the tangent-point circle, where $s=0$. Thus, the tangent-point circle is a singular region in this method. 

These  behaviors are just in the opposite sense to the case for equation \ref{eq-dvrad} and figure \ref{fig-vrad}. In this context  the radial-velocity method, including the tangent-point (terminal-velocity) method, and the proper-motion method are complimentary to each other, in so far as circular rotation assumption is made.

\subsection{Rotation Velocity $\Vrotv$ from Velocity Vector, and Accuracy Diagram $\Delta \Vrotv(X,Y)$}

If the radial velocity and proper motion as well as the distance are known at the same time for the same object, its three-dimensional velocity vector is determined without assuming circular orbit. The absolute value of the velocity vector is calculated by
\be
V=\sqrt{\Up^2+ \Ur^2},
\label{eq-Vvpvr}
\ee
where
\be
\Up=r \mu+\V0 \cosl
\ee
 and
 \be
 \Ur=\vr + \V0 \sinl.
\ee

Since the deviation of velocity vector from the circular orbit may be assumed to be small, we here neglect the error arising from the non-circular components, and define the rotation velocity by the velocity given here as $\Vrotv=V$. We now calculate the error included in the absolute value of the velocity vector by
\be
\Delta \Vrotv= \left[
\left({\partial V \over \partial \mu} \right)^2\delta \mu^2  
+\left({\partial V \over \partial \vr} \right)^2  \delta \vr^2
+\left({\partial V \over \partial r}\right)^2  \delta r ^2
\right]^{1/2}
\ee
\be
={1 \over V} \left[\Up^2 r^2 \delta \mu^2+\Ur^2 \delta \vr^2 + \Up^2 \mu^2 \delta r^2 \right]^{1/2}.
\label{eq-dvvec}
 \ee

 Figure \ref{fig-vvec} shows an accuracy diagram for velocity vector, or the distribution of $\Delta \Vrotv$ calculated for $\delta \vr=1$ \kms, $\delta \mu=0.21 $ mas y$^{-1}$, and $\delta r/r=0.02 $. This diagram shows a milder variation of error in $V$ near the tangent-point circle compared with that calculated for the radial velocity method in figures \ref{fig-vrad} and that for proper motion method in \ref{fig-vprop}. Also, figure \ref{fig-vvec}  shows milder error variations around the Sun-GC line. Thus, we see that the velocity-vector method has no singular regions to determine the rotation velocity, and provides us with more general information from the entire galactic disk.

\begin{table*}
\caption{Distances, proper motions, and radial velocities for star forming regions observed with VERA. }
\bc
\begin{tabular}{llllllllllll}
\hline\hline 
Source& $l$ &$b$&$r \pm \delta r$&$\vr \pm \delta \vr$ & $\mu \pm \delta \mu^\dagger$ &Reference\\
&(deg)&(deg)&(kpc) &(\kms) & (mas y$^{-1}$)\\
\hline 
IRAS 06058+2138 &188.9&0.88 &$1.76\pm 0.11$ & $3\pm 3$ &$2.57\pm 0.40$& (1)\\
IRAS 19213+1723&52.10&1.04 &$3.98\pm 0.57$&$41.7\pm 2$&$-14.66\pm 4.41$&(1) \\
AFGL 2789&94.60& -1.79& $3.07\pm 0.29$ & $-44\pm 2$ &$-4.03\pm 0.40$&(1)\\
G48.61+0.02&48.61& 0.02& $5.03\pm 0.19$& $ 19\pm 1$&$ -5.86\pm 0.20$&(2)\\  
ON1& 69.54& -0.98& $2.47\pm  0.11$&  $12\pm 1$ & $-7.50\pm 0.01$ &(3)\\  
ON2N & 75.78& -0.34& $3.83\pm 0.13$ & $0\pm 1$& $ -7.98\pm 0.03$&(4)\\    
\hline  \\
\end{tabular}\\
{\small (1) Oh et al. (2009); (2) Nagayama et al. (2011a); (3) Nagayama et al. (2011b); (4) Ando et al. (2011) }
\\
$\dagger$ Recalculated from $\vp$ given in the original papers.
\\ 
\label{tab-veradata} 
\ec

\caption{Values of $\Vrotr$, $\Vrotp$ and $\Vrotv$ and their errors for sources in table \ref{tab-veradata}. }
\bc
\begin{tabular}{llllllllllll}
\hline\hline 
Source& $R \pm \delta R$& $\Vrotr \pm \Delta \Vrotr$ & ${\Vrotp} \pm \Delta \Vrotp$&$\Vrotv \pm \Delta \Vrotv$ \\ 
&(kpc)&(\kms) & (\kms) & (\kms)& \\
\hline 
IRAS 06058&$9.74\pm 0.11$ &$219.9\pm 23.9$ &$177.6\pm 3.6$ &$178.4\pm 3.6$\\
IRAS 19213&$ 7.05\pm 0.25$& $223.0\pm 12.9$ &--- $^\ddagger$ &$199.5\pm 2.0$&\\
AFGL 2789 &$8.8-\pm 0.12$& $171.4\pm 16.7$ & $177.1\pm 16.7$ &$172.4\pm 4.0$ \\
G48.61&$6.01\pm 0.01$ & $169.2\pm 1.7$ & ---  & $169.2\pm 0.9$ \\
ON1& $7.50\pm 0.01$ & $199.6\pm 1.0$ & ---  & $199.4\pm 0.10$ &\\
ON2N & $7.98\pm 0.03$ & $199.4\pm 3.1$ & --- & $200.0\pm 1.5$ &\\
\hline 
\end{tabular}\\
$\ddagger$ Near tangent point.\\
\label{tab-veravrot} 
\ec
\end{table*}

\subsection{Observed examples}

Recently, trigonometric parallax and proper motion measurements of maser sources have obtained reliable data for galactic rotation determinations (Oh et al. 2009; Ando et al. 2011; Nagayama et al. 2011a, b). Combined with radial velocities from associated CO line, the data are available to calculate the rotation velocity. Table \ref{tab-veradata} shows the observed parameters from the VERA observations, where the errors for systemic CO velocity were read from the original papers cited therein. We calculate $\Vrotr$, $\Vrotp$, and $\Vrotv$ and their errors from the data in this table. Table \ref{tab-veravrot} shows the calculated results.

The thus derived values for rotation velocities are consistent with each other as well as with those derived by Oh et al. (2009). However, some peculiar results are obtained in proper-motion method for near tangent-point sources, whose values are not listed. In general, the velocity-vector method gives more reliable results. The slight differences in the here derived rotation velocities $\Vrotv$ and those by Oh et al. is due to neglecting motions perpendicular to the galactic disk in this paper, since we aimed at examining the accuracy diagrams within the galactic disk.

\section{Radial-Velocity and Proper-Motion Fields and Kinematical Distances}

Once the rotation curve is determined, and if we assume circular rotation, the rotation curve may be in turn used to measure kinematical distances of objects by applying the velocity-space transformation (Oort et al. 1958; Nakanishi and Sofue 2003, 2006). The kinematical distance is obtained either from radial velocity or from proper motion using equations (\ref{eq-vr}) and (\ref{eq-vp}).  

\subsection{Radial-Velocity Field $\vr(X,Y)$ }

If we assume circular motion, the velocity field can be used to derive kinematical distance $r_{\vr}$ by measuring the radial velocity $\vr$. This velocity-to-space transformation is useful to map the density distribution of interstellar gases from HI and/or CO emission lines (e.g. Nakanishi and Sofue 2003). 
The kinematical distance $r$ is given by
\be
r=R_0 \cosl \pm \sqrt{R^2-R_0^2 {\rm sin}^2 l}
\label{eq-r}.
\ee
Here the galacto-centric distance $R$ is related to the radial velocity through equation \ref{eq-vrot}, where $V_{\rm rot}^{\vr}$ is replaced with the determined rotation velocity $V(R)$. Then, we obtain 
\be
R=R_0 V(R) \left({\vr \over  \sinl}+V_0 \right)^{-1}.
\label{eq-R}
\ee

\begin{figure}
\bc
\includegraphics[width=7.5cm]{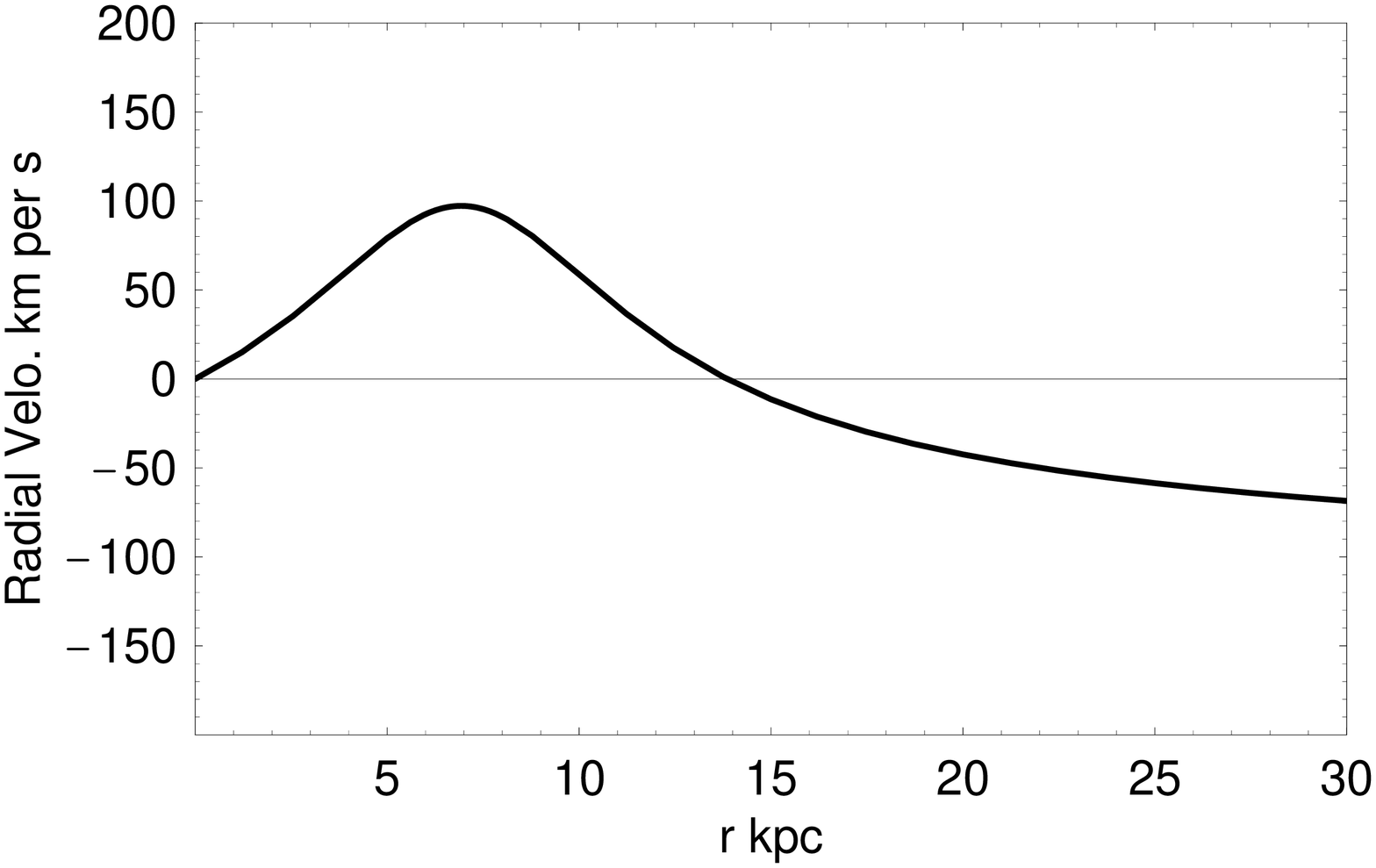}  \\ 
\ec
\caption{Variation of radial-velocity $\vr$ as a function of the line of sight distance $r$ at $l=30^\circ$. }
\label{fig-vr-r}

\bc
\includegraphics[width=7.5cm]{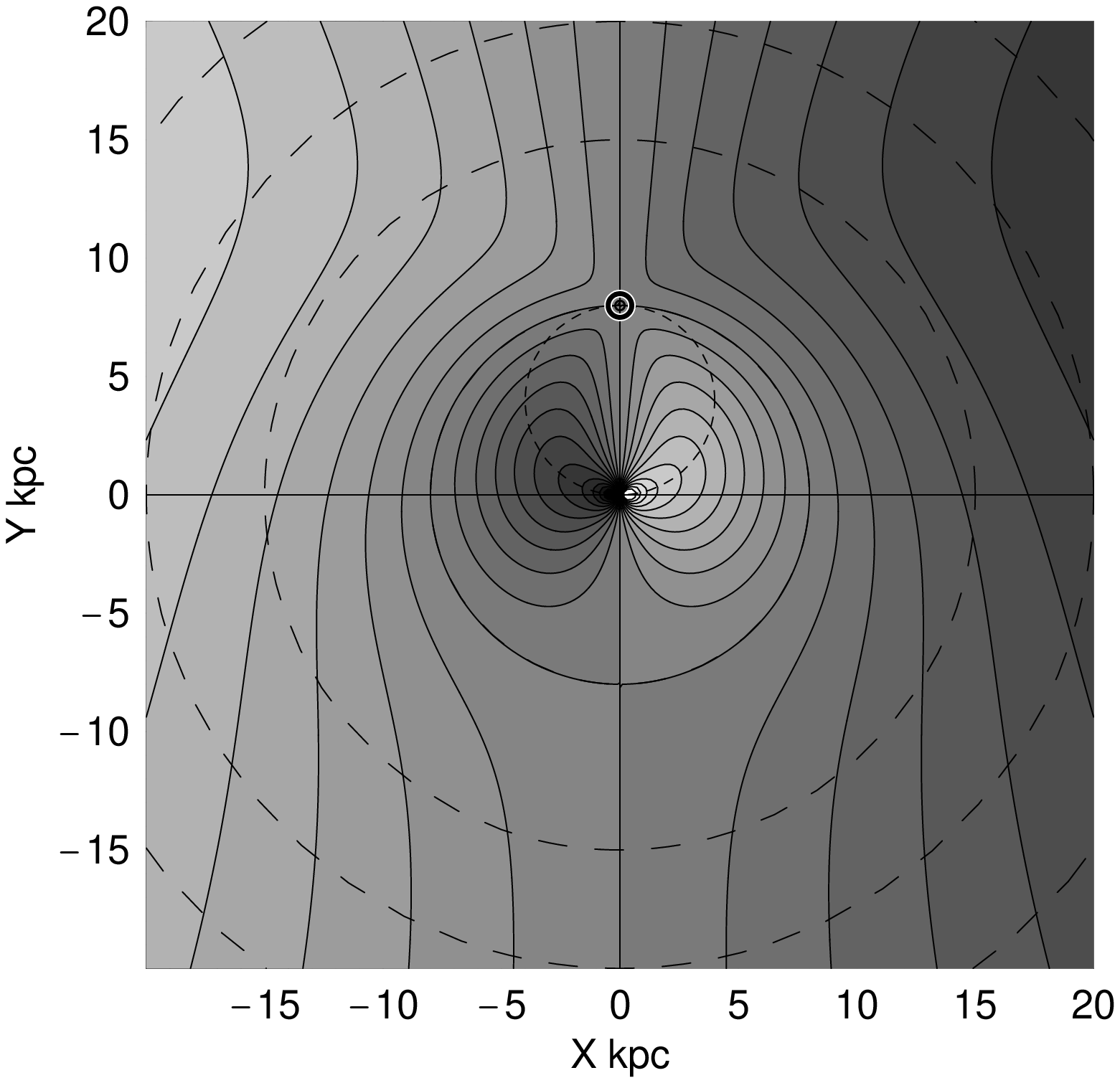}  \\ 
\ec
\caption{Radial-velocity  field, $\vr(X,Y)$, for the rotation curve in figure \ref{fig-rcmodel}. Contours are drawn every 20 \kms interval with bright region being positive and dark being negative velocities.  }
\label{fig-vfi}
\bc
\includegraphics[width=7.5cm]{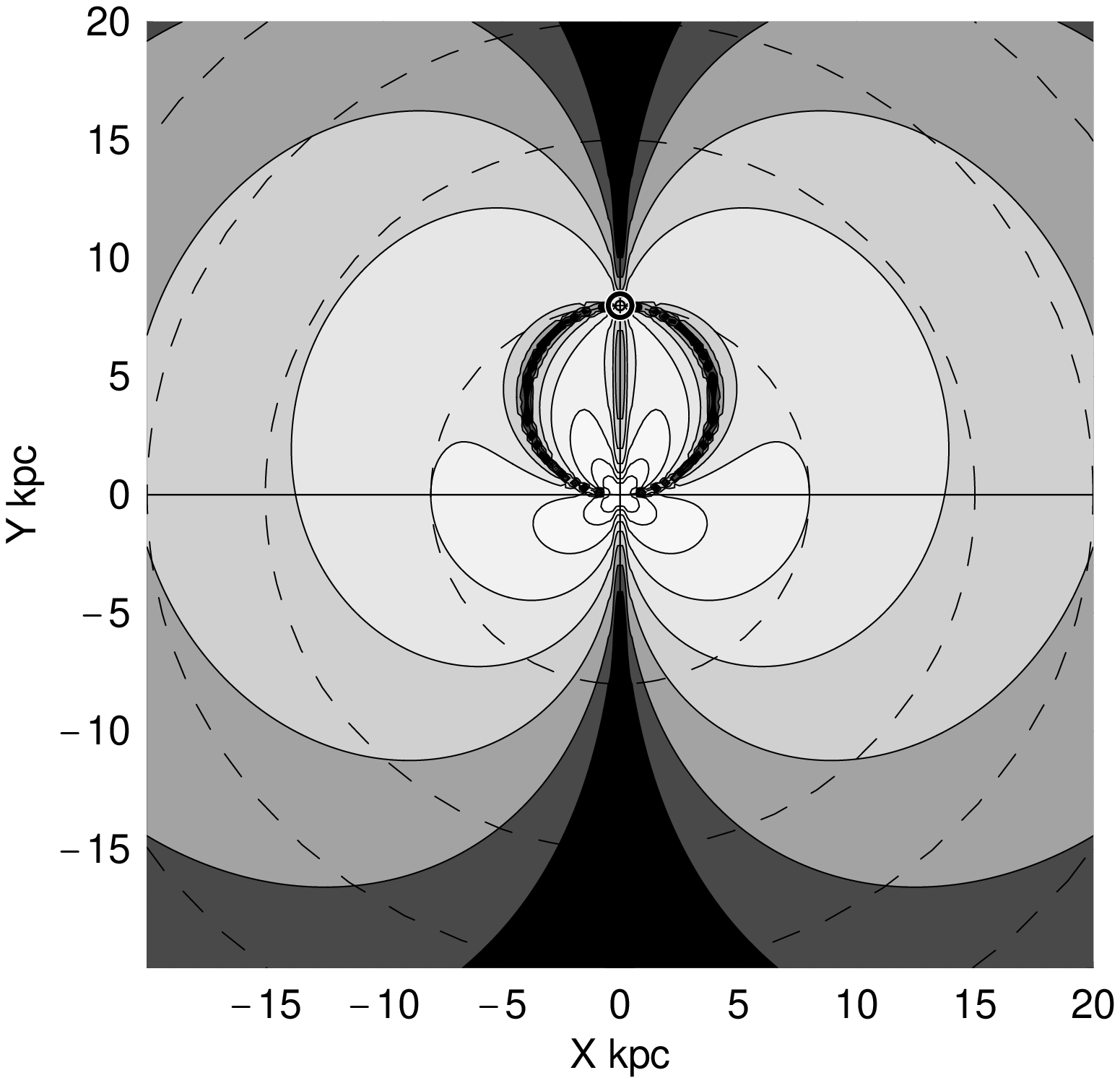}  \\ 
\ec
\caption{Accuracy diagram,  $\Delta r_{\vr} (X,Y)$, for $\delta \vr=1$ \kms. Contours are drawn at $\delta r=$0.01, 0.02, 0.04, 0.08, 0.16, 0.32, 0.64, .... kpc from white to dark.}
\label{fig-adr}
\end{figure}

Figure \ref{fig-vr-r} shows the variation of $\vr$ as a function of $r$ at $l=30\circ$. Figure \ref{fig-vfi} (a) shows a radial velocity field, e.g. the distribution of $\vr$ on the galactic plane, calculated for the model rotation curve in figure \ref{fig-rcmodel}. Such velocity fields have been often used to obtain the distribution of interstellar gases from radial velocities of the HI line and CO molecular lines (e.g. Nakanishi and Sofue 2003; 2006). 

\subsection{Accuracy diagram  $\Delta r_{\vr}(X,Y)$}

The accuracy of the velocity-to-space transformation depends on the accuracy of the kinematical distance $r_{\vr}$. We now construct an accuracy diagram for $r_{\vr}$ by differentiating equation (\ref{eq-r}) with respect to $\vr$ to obtain the error $\Delta r_{\vr}$ of $r_{\vr}$ in terms of the error $\delta \vr$ of $\vr$. Since $V(R)$ is a slow function of $R$ and $r$, we may neglect the terms including the derivative of $\partial V(R)/\partial r ~\delta r$. Note that this approximation does not hold in the Galactic Center at $R<\sim 0.5$ kpc, so that the following results is not precise enough in the central region. We, then, obtain
\be
\Delta r_{\vr} ={R^3 \over R_0} 
{1 \over \sqrt{ {\rm sin}^2 l (R^2- R_0^2  {\rm sin}^2 l ) } }
{\delta \vr \over V(R)}
\label{eq-dr}
\ee

Figure \ref{fig-adr} shows the accuracy diagram of the kinematical distance $r$, or the distribution of $\Delta r_{\vr}$ on the galactic plane, calculated for $\delta \vr=1$ \kms using the model rotation curve shown in figure \ref{fig-rcmodel}. It is trivial that the distance error is largest along the Sun-GC line, where the motion by galactic rotation is perpendicular to the line of sight. The figure shows that the tangent-point circle is a singular region, where the distance determination cannot be applied. This is deeply related to the near-far ambiguity problem in solving the kinematical distance  inside the solar circle.

\subsection{Proper-Motion Field $\mu(X,Y)$}

Given a rotation curve $V(R)$, and if we assume circular rotation, the proper motion of an object is given by
\be
\mu=-{1 \over r}\left({s \over R}V(R) +V_0  \cosl \right).
\label{eq-mu}
\ee
Figure \ref{fig-mu-r} shows the variation of $\mu$ as a function of distance $r$ in the direction of $l=30^\circ$. Figure \ref{fig-mufield} shows the $\mu$ field, or the distribution of proper-motion  on the galactic plane, calculated for the model rotation curve in figure \ref{fig-rcmodel}. Obviously, an object on the solar circle has proper motion of 
\be
\mu_\odot=-{V_0 \over R_0} =\Omega_0 .
\ee
For our present values $R_0=8$ kpc and $V_0=$ 200 \kms, we have $\mu_{\rm \odot}=-5.26$ mas y$^{-1}$. 

This kind of $\mu$ field may have not been used in the current studies for galactic dynamics because of the lack in a sufficient number of objects with measured proper motions. However, the progress in VLBI trigonometric measurements have made it possible to apply such a diagram for determination of kinematical distance $r_\mu$. It must be noted that there is no singular region beyond the Galactic Center.

Remembering $R=\sqrt{r^2+R_0^2-2rR_0 \cosl}$ and $s=r-R_0 \cosl$, we may iteratively solve the above equations to obtain the kinematical distance $r$ in terms of $\mu$. We first start with an arbitrary initial value of $r=r_1$ to calculate the corresponding proper motion $\mu_1$. 
Then the difference $\delta \mu=\mu_{\rm ob}-\mu_1$ is related to a correction $\delta r$ to $r$ as
\be
\delta r_\mu={- r \delta \mu \over \mu + (1/R - s^2/R^3)V(R)}.
\label{eq-muiter}
\ee
Now, the initial value of $r_\mu=r_1$ is corrected for thus obtained $\delta r_1$ to yield the second approximate value $r_2=r_1+\delta r_1$.
Equation (\ref{eq-mu}) is then used to find the next approximate value $\mu_2$, which is further used to get the second correction $\delta r_2$ to obtain $r_3$ and $\mu_3$. In this way, the iteration may be repeated until the difference $\delta \mu_{i+1}=\mu_{\rm ob}-\mu_i$ becomes sufficiently small compared to the observational error. Since the $\mu$ field has a mild variation over the galactic disk as shown in figure \ref{fig-mufield}, the iteration usually results in a sufficiently stable value within several times, e.g. within $i=4$ to 5. If the $\mu$ value exceeds the maximum value plotted in figure \ref{fig-mufield}, we have no solution, but the iteration diverges. 

Table \ref{tab-muiteration} shows an example of the convergence for a case of an assumed object at $l=30^\circ$ and $\mu=-4$ \masy, starting from an arbitrary distance at $r=15$ and 2 kpc corresponding to far and near-side solutions. The table shows the rapid convergence of the iteration for both sides.

\begin{table*}
\caption{Example of convergence of $\mu$ and $r_\mu$ for an assumed object for far and near side solutions.}
\bc
\begin{tabular}{llllllllllll}
\hline\hline 
  & & Far solution &  &Near solution& \\
$i$  &$l$& $\mu$ &  $r_\mu$ & $\mu$&  $r_\mu$\\
& (deg)& (\masy) & (kpc) & (\masy)&  (kpc)\\
\hline 
0 &$30^\circ$& $-4.0 \pm 0.1$  &15.0 &  $-4.0 \pm 0.1$&2.0\\
1 &&-4.93949  &18.5053& -1.83192 &7.34575\\
2 && -4.09845 &18.993 & -5.55054 &4.90354\\
3 && -3.99913 &18.9885& -3.57916 &5.43869\\
4 && -4.00003 &18.9886& -4.02638 &5.40723\\
5 && -4.      &18.9886& -3.99951& 5.40782\\ 
6 && -4.      &$18.9886 \pm 0.44$ & -4.00001& $5.40781\pm 0.043 $ \\ 
\hline  \\
\end{tabular}
\\
\label{tab-muiteration}
\ec

\caption{Proper-motion distances, $r_\mu$, determined for the VERA data in table \ref{tab-veradata}.}
\bc
\begin{tabular}{llllllllllll}
\hline\hline 
Source& $l$ & $r_\pi \pm \delta r_\pi$  & $\mu \pm \delta \mu$ & $r_\mu \pm \delta r_\mu$ \\
&(deg)&(kpc) &(mas)&(kpc)\\
\hline 
IR06058  &188.9 & $1.76\pm 0.11$ & $ 2.6 \pm 0.4$ &  --- (no solution)\\
IR19213  &52.10 & $3.98\pm 0.57$ & $ -6.5 \pm 0.8$ & --- \\
AFGL2789        &94.60 & $3.07\pm 0.29$ & $ -4.0  \pm 0.5 $ & $5.5 \pm 0.4$\\
G48.61      &48.61 & $5.03\pm 0.19$ & $ -5.8 \pm 0.1$ & $6.5\pm 0.1$ \\  
ON1              &69.54 & $2.47\pm  0.11$& $ -6.0 \pm 0.7$ & ---\\  
ON2N             &75.78 & $3.83\pm 0.13$ & $ -5.4 \pm 0.2$ & ---\\  
\hline  \\
\end{tabular}
\label{tab-mudistance}
\ec
\end{table*}

\subsection{Accuracy diagram $\Delta r_\mu(X,Y)$}

The error propagation is estimated from equation (\ref{eq-mu}) by giving small perturbations to $r$ and $\mu$. We again neglect the term including  $(\partial V(R)/\partial r) \Delta r$. Then $\Delta r_\mu$, or the error in $r_\mu$, may be expressed in terms of $\delta \mu$ as
\be
\Delta r_\mu= {r^2 \delta \mu \over V(R) \left[r s^2 / R^3- (r+s)/ R \right]-V_0 \cosl}.
\label{eq-drmu}
\ee

\begin{figure}
\bc
\includegraphics[width=7.5cm]{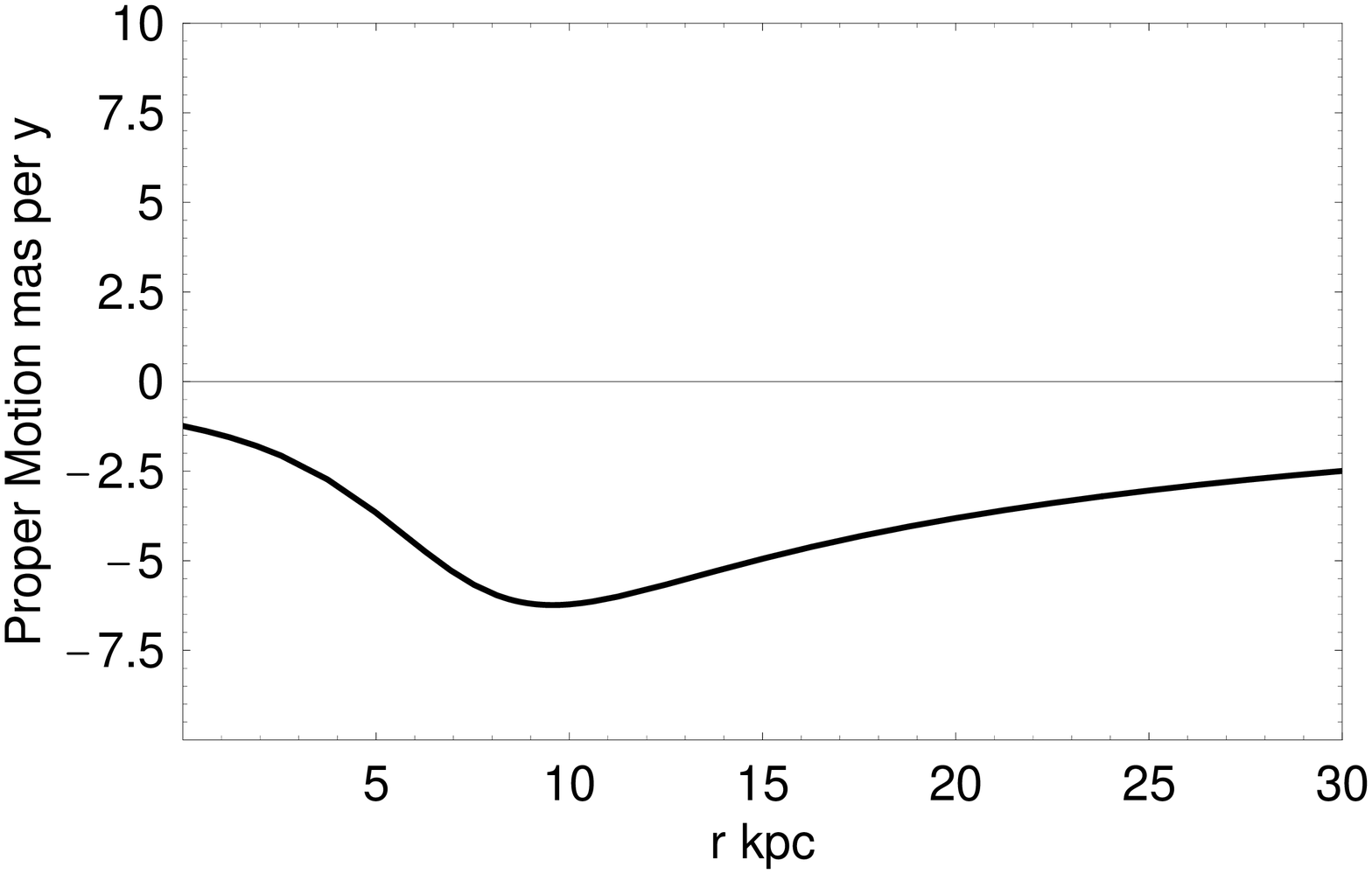}  \\ 
\ec
\caption{Variation of proper motion $\mu$ as a function of the line of sight distance $r$ at $l=30^\circ$.}
\label{fig-mu-r}

\bc
\includegraphics[width=7.5cm]{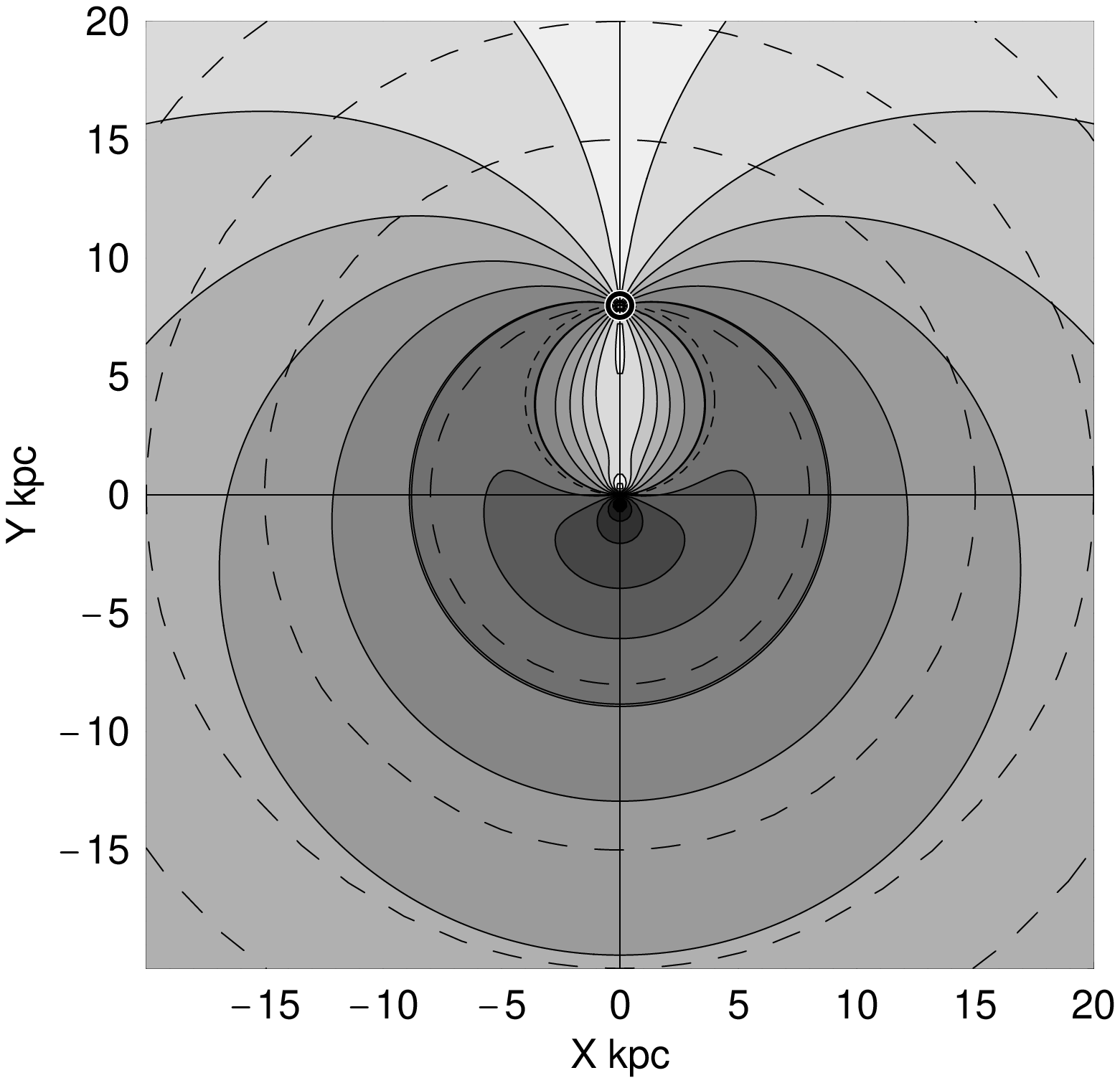}  \\ 
\ec 
\caption{Proper-motion field, $\mu(X,Y)$, for the rotation curve in figure \ref{fig-rcmodel}. Contours are drawn every 1 mas y$^{-1}$ with the thick line near the solar circle being -5 mas y$^{-1}$. }
\label{fig-mufield}  
\bc
\includegraphics[width=7.5cm]{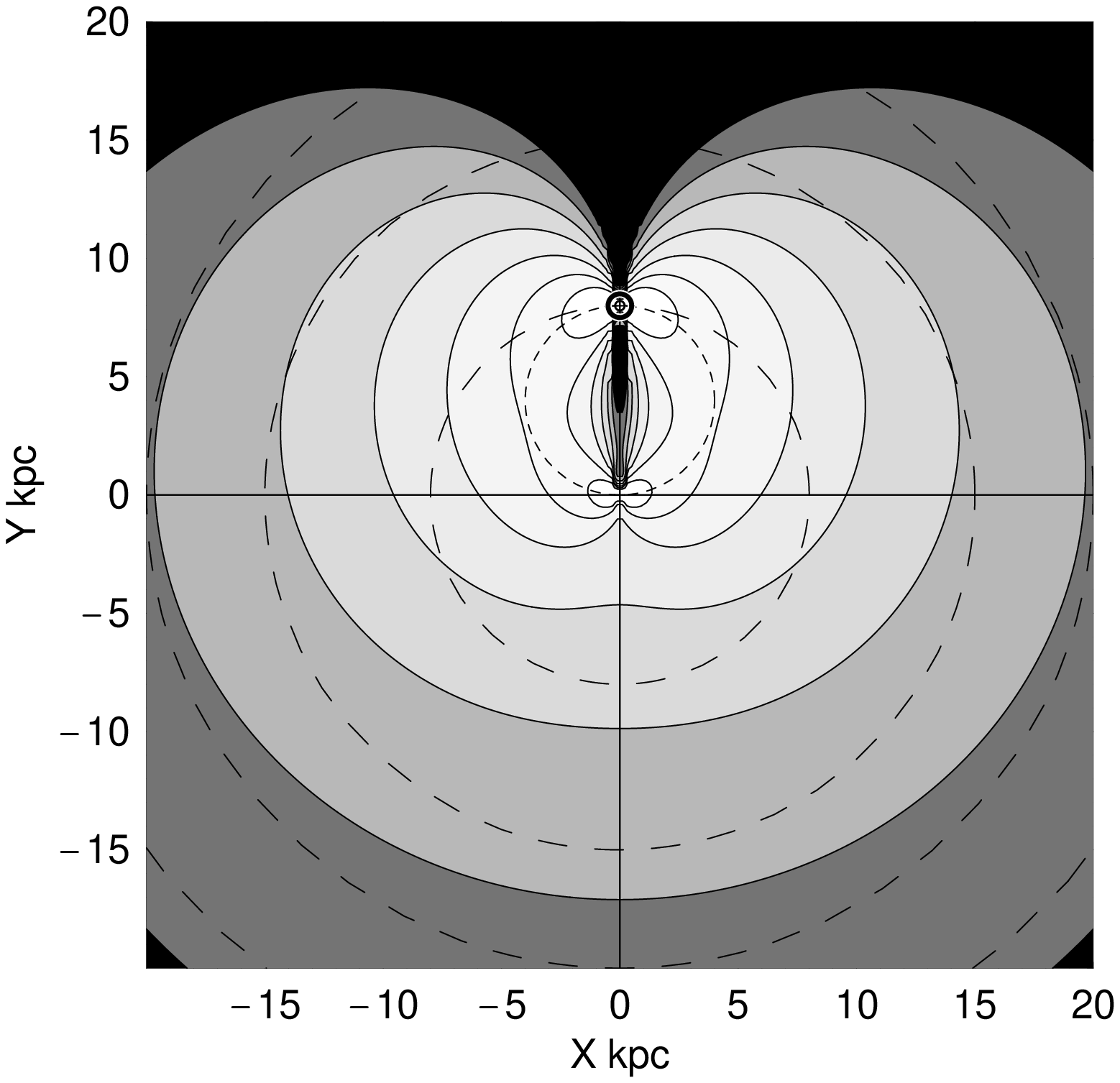}  \\ 
\ec 
\caption{Accuracy diagram, $\Delta r_\mu(X,Y)$, for $\delta \mu =0.2$ \masy.  Contours are drawn at $\delta r=$0.01, 0.02, 0.04, 0.08, 0.16, 0.32, 0.64, .... kpc from white to dark.}
\label{fig-admufi}  
\end{figure}

Figure \ref{fig-admufi} shows the accuracy diagram for kinematical $\mu$ distance $r_\mu$, or the distribution of $\Delta r_\mu$ on the galactic plane, calculated for $\delta \mu=0.2$ \masy. The figure indicates that the distance ambiguity is largest along the Sun-GC line in the near side of the Galactic Center, where the objects have proper motions $\mu \sim \left[V(R)-V_0\right]/r$, which is usually small because of the flat rotation curve. 

On the other hand, the ambiguity is drastically reduced in the region beyond the Galactic Center, where the object moves in the opposite direction to the Solar motion,  perpendicularly to the line of sight at about twice the rotation velocity, yielding large proper motion 
\be
\mu \sim -{V(R)+V_0 \over r}, 
\ee
yielding $\mu=-5.26$ \masy for a solar circle object beyond the Galactic Center.
Hence, the distances of Sun-GC line objects beyond GC may be determined with relatively good accuracy from the proper motion method. This is particularly important, because distances cannot be measured by radial-velocities. Also, the trigonometric (parallax) measurements would be still difficult for such distant objects beyond GC, whereas the proper motion is sufficiently large.

We here try to apply the proper-motion method for kinematical distance $r_\mu$ to the observed $\mu$ values given in table \ref{tab-veradata}. Table \ref{tab-mudistance} lists the thus derived distances. We obtained converged solutions for the two sources, FGL2789 and G48.61+0.02. The iteration diverged for the other sources, whose proper motions exceed the expected values from the model rotation curve. This may arise either due to intrinsically non-circular motions of the sources (e.g. Oh et al. 2009), or by the adopted rotation curve which might not well represent the true galactic rotation.

\section{Discussion}

We have analyzed the expected errors in rotation curve determinations arising from observational errors of distance, radial velocity and proper motion for individual objects depending on their galactic positions. We displayed the error distributions for derived rotation velocities from the various methods as accuracy diagrams. We presented the accuracy diagrams for some combinations of assumed errors in $r$, $v_r$ and $v_p$, while they represent the general properties of the error distributions. If we adopt different combinations of intrinsic observation errors, the diagrams will change quantitatively, but their general characteristics are similar to those presented here.

The radial velocity method assuming circular motion has been most often used in the decades. The tangent point method for the inner rotation curve using the HI and CO line emissions is an extreme case choosing objects on the loci of the minimum errors in figure \ref{fig-vrad}. The accuracy diagram, $\Delta \Vrotr(X,Y)$, well explains the reason why the observed rotation curve in figure \ref{fig-rc} is nicely determined at $R<8$ kpc by the tangent-point method compared to the outer rotation curve. The tangent-point circle is a special region where the radial velocity method can give the highest accuracy rotation curve. Furthermore, this diagram suggests that the butterfly areas at $l\sim 100-135\deg$ and $l\sim 225-280\deg$ are suitable regions for selecting the sources for determination of outer rotation curve in this method. It should be mentioned that similar accuracy is expected for sources within the tangent-point circle in order to determine the inner rotation curve. Sources near the Sun-GC line are, of course, not appropriate for this method, as the accuracy diagram shows singularity.

In the proper motion method assuming circular motion, the most accurate measurement of rotation velocity is obtained for objects near the Sun-GC line as shown by the accuracy diagram, $\Delta \Vrotp(X,Y)$, in figure \ref{fig-vprop}, as indeed realized by Honma et al. (2007). It must be also emphasized that the minimum error area is widely spread over $l \sim 120 - 250\deg$ in the anti-center region, as well as in the central region inside the tangent-point circle. The largest error occurs for objects lying near the tangent-point circle. Thus, the tangent-point circle is singularity circle in this method.

In the 3-D velocity vector measurement, where no assumption of circular motion is made, the three independent values of the radial velocity, proper motion and distance are required at the same time. The accuracy diagram, $\Delta \Vrotv(X,Y)$, in figures \ref{fig-vvec} indicates milder dependence of the accuracy on the location of the sources compared to those in the previous two methods. Optimization of source selection is, therefore, easier in the 3-D method. 

Analyses of the accuracy diagrams presented in this paper may summarize the behaviors observed in the galactic rotation curves obtained in the decades. In the current observations, source selection has  been optimized by the authors a priori, while it was not necessarily in a systematic way. The present analysis may be helpful for further source selections in the future in order to optimize the observations for higher accuracy rotation curves from limited resources of observing time and facilities.

Once a rotation curve is obtained, it may be in turn used to map the ISM and stellar objects in the galactic plane from their radial velocities and proper motions. The former method has been often applied to map the Galaxy. The latter method, however, is not used often because of the lack of proper motion measurements. The recent VLBI trigonometric measurements of proper motions would be challenging in mapping the Galaxy. We have presented the radial velocity field, $\vr(X,Y)$, and proper motion field, $\mu(X,Y)$, for an assumed circular rotation curve as in figure \ref{fig-rcmodel}. We constructed accuracy diagrams, $\Delta r_{\vr}(X,Y)$ and $\Delta r_\mu(X,Y)$,  for distance measurements using the velocity and proper motion fields. The $\Delta r_{\vr}(X,Y)$ field confirms the current behaviors seen in the galactic maps so far published in the decades. The $\Delta r_\mu(X,Y)$ also confirms that the proper motion measurements are powerful tool to solve the distance ambiguity in the Galaxy along the Sun-GC line, particular for sources beyond the Galactic Center. 
 
Finally, we comment on possible systematic errors. Non-circular motions by the bar, spiral arms, random motions, tidal effects by the companion galaxies, etc.., would be superposed on the rotation curve. Such effects will cause systematic errors in the determined quantities, and affect the error analyses. The uncertainties in the adopted galactic constants like $R_0$ and $V_0$ affect the results, which, moreover, affect the adopted rotation curve to calculate the errors. Hence, the present error analyses include circularity among the evaluated quantities and errors. This circularity means that the results should not be interpreted too rigorously, but are to be used as a guide to the galactic dynamics based on the adopted galactic constants and the assumption of circular rotation.

{}

\end{document}